\documentstyle[eqsecnum,aps,psfig,12pt]{revtex}

\def\step{\\[3ex]}

\def\ov{\over}
\def\CO{{\mathcal O}}
\date{\today}
\let\Ga=\Gamma

\def\ds{\displaystyle}

\def\dcal{{\cal D}}

\def\tr{{\rm tr}}
\def\Tr{{\rm Tr}}
\def\Der{{\bf {\cal{G}}}}
\def\Loop{{L}}
\def\eq#1{(\ref{#1})}
\def\Eq#1{Eq.~(\ref{#1})}
\def\di{\displaystyle}
\def\longbar#1{#1\kern-0.7em\raise1.3ex\hbox{{$-$}}}
\def\tab{&\di}
\let\de=\delta

\begin{document}
\thispagestyle{empty}
\begin{flushright}
{\tt hep-th/0009110\\DIAS-STP-00-17\\
 HD-THEP-00-45}\,\,\, 
\end{flushright}
\vspace{1cm}
\begin{center}
{\large \bf Gauge invariance and background field formalism}\\[1ex]
{\large \bf in the exact renormalisation group}\\[9.5ex]
 
{Filipe Freire,} ${}^{a,b}$\footnote{freire@thphys.may.ie}
{} 
{Daniel F. Litim} ${}^{c}$\footnote{D.Litim@thphys.uni-heidelberg.de}
{and}
{Jan M. Pawlowski} ${}^{b}$\footnote{jmp@stp.dias.ie}
\\[2ex]

${}^a${\it Department of Mathematical Physics, N.U.I. Maynooth, Ireland.\\[2ex]

       ${}^b$School of Theoretical Physics,
       Dublin Institute for Advanced Studies,\\
       10 Burlington Road, Dublin 4, Ireland.}\\[2ex]

${}^c${\it  Institut f\"ur Theoretische Physik, Philosophenweg 16,\\
  D-69120 Heidelberg, Germany.}\\[6ex]

{\small \bf Abstract}\\[2ex]
\begin{minipage}{14cm}{\small
We discuss gauge symmetry and Ward-Takahashi identities for 
Wilsonian flows in pure Yang-Mills theories. The background
field formalism is used for the construction of a gauge invariant 
effective action. The symmetries of the effective action under gauge 
transformations for both
 the gauge field and the auxiliary background field are 
separately evaluated. We 
 examine how the symmetry properties of the full 
theory are restored in the limit where the cut-off is removed. \\[5ex]

}
\end{minipage}
\end{center}
\newpage
\pagestyle{plain}
\setcounter{page}{1}

\renewcommand{\thefootnote}{\arabic{footnote}}
\setcounter{footnote}{0}

\noindent
{\bf 1. Introduction}\\[-1ex]

The Wilsonian or exact renormalisation group (ERG) \cite{Polchinski:1984gv}
has been successfully applied to both perturbative and
non-perturbative phenomena in field theory. The main advantages of
such an approach are its flexibility and the comparatively simple
numerical implementation. Applications to gauge theories are much more
involved because it is less obvious how a Wilsonian cut-off can be
implemented for a (non-linear) gauge symmetry. Much work has been 
devoted to overcoming this intricacy 
\cite{Reuter:1994kw,Ellwanger:1994iz,Ellwanger:1998wy,Bonini:1994sj,Freire:1996db,D'Attanasio:1996jd,Litim:1998qi,Morris:2000px,Litim:1998nf} 
(see \cite{Litim:1998nf} for a review). We 
focus the discussion on 
pure Yang-Mills theory, since the inclusion of fermions is straightforward. 
The ERG equation for the corresponding effective
action $\Gamma_k$ describes how $\Gamma_k$ changes under an
infinitesimal variation of the infra-red scale $k$: 
\begin{eqnarray} 
\partial_t\Gamma_k[A,c,c^*;\bar A] =
\frac{1}{2}\Tr\left(\frac{\de^2\Gamma_k}{\delta A\delta A}
              +R_{\!A}\right)^{\!\!\!-1}\!\!\!\!\partial_t  R_{\!A}
\,- \,
\Tr \left( \frac{\de^2\Gamma_k}{\delta c \delta c^*} 
    +R_{C}\right)^{\!\!\!-1}\!\!\!\! \partial_t
              R_{C}\,.
\label{flow}
\end{eqnarray}
Here, $t=\ln k$ is the logarithmic scale parameter and the trace $\Tr$ 
denotes a sum over momenta, Lorentz and gauge group indices. 
The functions $R_A$ and $R_{\hspace{-.01cm}C}$ implement the
Wilsonian infra-red cut-off for the gauge field $A$ and the
ghost fields $c$ and $c^*$ 
respectively.
We also introduced an non-dynamical auxiliary field $\bar A$, the so-called 
background gauge field. 

In the present Letter we discuss in detail the symmetry properties of
coarse-grained effective actions for non-Abelian gauge theories 
and their flows \eq{flow} within the background field approach. A similar 
programme has been put forward for Abelian gauge theories in 
\cite{Freire:1996db}. The
most attractive feature of a background field formalism is that it
provides a gauge invariant effective action \cite{Abbott:1981hw}, which is
defined via an identification of the auxiliary background field with
the original gauge field. This property can be  maintained
even within a Wilsonian approach \cite{Reuter:1994kw}. 

The symmetry properties of a background field effective action are
naturally encoded in Ward-Takahashi identities. These identities are
derived  by applying gauge transformations separately to  dynamical
fields and background field. 
Clearly, the investigation of the Ward-Takahashi identities play a
pivotal r$\hat{\rm o}$le in the evaluation of the coarse grained
effective action. We shall argue that 
it is crucial to separately discuss the
action of gauge transformations on the  dynamical fields and the gauge
transformations on the background field. Solving the related 
Ward-Takahashi (or BRST) identities poses a fine-tuning problem which is 
known to be soluble in perturbation theory. For Wilsonian flows these 
identities have been discussed in 
\cite{Ellwanger:1994iz,Bonini:1994sj,D'Attanasio:1996jd}. In the present 
approach we deal with an additional Ward-Takahashi identity related to 
gauge transformations of the background field. 
Whether this imposes an additional fine-tuning condition is an important 
question which has not yet been addressed. 

In the present contribution we close this conceptional gap in the formalism. 
The functional form of the Ward-Takahashi identities is established for
both transformations. We argue that no additional fine tuning
problem related to the existence of the new identity arises. Both
Ward-Takahashi identities are shown to be compatible with the 
flow. This guarantees that both the usual Ward-Takahashi identity and
the background field identity hold in the limit where the cut-off is
removed.  In this manner the gauge invariance displayed in the
effective action is the physical gauge invariance rather than an
auxiliary symmetry. \step

\noindent
{\bf 2. Background field formalism}\\[-1ex]

We briefly summarise some important points about the background
field formalism, in particular the r$\hat{\rm o}$le of the different
gauge transformations present in this approach. 
In the background field formalism an auxiliary (non-dynamical) gauge 
field $\bar A$ is introduced. The formalism then
hinges on the use of a gauge fixing condition which depends on 
this field in such a way that the condition is 
invariant under a simultaneous gauge transformation of $\bar A$ 
and of the dynamical fields $A, c$ and $c^*$. This can be used to
 define an effective action which is invariant under 
the combined gauge transformation mentioned above. 
As the auxiliary field $\bar A$ is involved in 
this transformation it is clear that 
the invariance of the effective action is, {\it a priori}, 
an auxiliary symmetry. The essential point is that this symmetry for
the special choice $\bar A=A$ becomes the inherent gauge symmetry 
of the theory. 

The starting point is the
gauge-fixed classical action for a pure Yang-Mills theory including 
the ghost term 
\begin{eqnarray}\label{S} 
S=S_{\rm cl} + S_{\rm gf} + S_{\rm gh}\,.
\end{eqnarray} 
The classical action $S_{\rm cl}=\frac{1}{4}\int_x
F^a_{\mu\nu}F^a_{\mu\nu}$ contains the field strength tensor
$F_{\mu\nu}(A) = \partial_{\mu} A_\mu -\partial_\nu A_{\mu} +
g\,[A_{\mu},\,A_\nu]$, where $F_{\mu\nu} \equiv F^a_{\mu\nu}t^a$ and
$A_{\mu} = A^a_{\mu} t^a$ with the generators $t^a$
satisfying $[t^a,\,t^b]=f^{abc}t^c$ and $\tr\, t^a t^b 
=-\frac{1}{2}\,\de^{ab}$. We also employ the
shorthand notation $\int_x \equiv \int d^dx$. In the adjoint
representation, the covariant derivative is  
\begin{eqnarray}\label{covD}
D_\mu^{ab}(A) = \delta^{ab}\partial_\mu + g f^{acb}A^c_\mu\ .  
\end{eqnarray} 
The natural choice for the gauge fixing is the background field gauge 
\begin{equation}\label{gauge} 
S_{\rm gf} =-\frac{1}{2\xi}\int_x\ 
(A-\bar A)^a_{\mu}\,\longbar D{}_\mu^{ab} \longbar D{}_\nu^{bc}\, (A-\bar A)^c_{\nu}
\end{equation}
which involves the covariant derivative 
$\longbar D{}\equiv D(\bar A)$. The corresponding ghost action is given by 
\begin{equation}\label{ghost} 
S_{\rm gh} =-\int_x c^*_a \,\longbar D{}_\mu^{ac}D_\mu^{cd}\,c_d \ .
\end{equation}
We now turn to the symmetries of the action in \eq{S} 
and introduce two different gauge
transformations. The first one, given by the infinitesimal generator $\Der^a$, 
gauge transforms the original set of fields $A,\,c,\,\textrm{and}\,c^*$.
It generates gauge transformations representing the underlying symmetry 
of the theory. 
The transformation is defined on arbitrary functionals of $A,\,c,\,c^*$ and
$\bar A$ as
\begin{eqnarray}\label{delkernel} 
\Der^a\tab = \tab D_\mu^{ab}\frac{\delta}{\delta A_\mu^b} -
g\,f^{abc}\left( c_c\, \frac{\delta }{\delta c_b}+
  c^*_c\,\frac{\delta }{\delta c^*_b}\right). 
\end{eqnarray}
{\it Finite} gauge transformations with parameter $\omega^a$ are generated 
by $\exp[-i\int_x\omega^a \Der^a]$. The action of \eq{delkernel} on the fields 
is given by  
\begin{eqnarray}
\Der^a(x) A^b_{\mu}(y) =   D^{ab}_{\mu,x}(A)\delta(x-y), \tab  
\qquad\quad \tab \Der^a(x)  c_b(y)    = -   g  f^a{}_{bc}\, c_c(x)\delta(x-y),
\nonumber \\[1ex]\di 
\Der^a(x) \bar A^b_{\mu}(y)  =  0,\ \qquad\qquad \qquad \tab \quad \tab 
 \Der^a(x) c_b^*(y)  =  -  g  f^a{}_{bc}\, c^*_c(x)\delta(x-y).
\label{del}
\end{eqnarray} 
The gauge field $A$ is transformed inhomogeneously, the ghosts transform 
as tensors according to their representation and the background field
is invariant. The subscript $x$ for the covariant derivative refers
to the variable on which the derivative operates and it will be
omitted whenever it is unambiguous.
From~(\ref{del}) it follows that the covariant derivative transforms
as a tensor,
\begin{eqnarray}
  \label{delcov}
  \Der^a(x)D^{bc}_{\mu,y}=g f^{bdc} D_{\mu,x}^{ad}\delta(x-y)\equiv 
 g\,\left([t^a\,\delta(x-y),\,D_{\mu,y}]\,\right)^{bc}.
\end{eqnarray}
The second gauge transformation, given by the generator 
$\bar\Der{}^a$, transforms only the background field 
$\bar A$
\begin{eqnarray}\label{delbarkernel} 
\bar\Der^a\tab =\tab \longbar D{}_\mu^{ab}\frac{\delta}{\delta \bar A_\mu^b}. 
\end{eqnarray}
On the fields it acts as
\begin{eqnarray}
\bar\Der^a(x) \bar A^b_{\mu}(y) \tab = \tab 
\longbar D{}^{ab}_{\mu}\delta(x-y), \qquad \qquad
\bar\Der^a A^b_{\mu} = \bar\Der^a c_b  
        = \,\bar\Der c_b^* = 0\ . 
\label{delbar}
\end{eqnarray}
Since $\bar\Der{}^a$ acts on $\bar A$ as $\Der{}^a$ on $A$ it 
follows that the covariant derivative $\longbar D$ transforms as a tensor as 
displayed in \eq{delcov} replacing $A$ with $\bar A$. 
$\bar\Der^a$ 
transforms the background field inhomogeneously while leaving the 
dynamical fields unchanged.
The background gauge transformation $\bar\Der^a$ is introduced as an auxiliary 
transformation which, as it stands, does not carry any physical information. 
We remark that \eq{delbar} implies $\bar\Der^a(x)
(A-\bar A)^b_\mu(y)=-\longbar D{}^{ab}_{\mu,x}\delta(x-y) $.

Let us now study the action of $\Der^a$ and $\bar\Der^a$ on the action $S$. 
The classical action is trivially invariant under both
the gauge symmetry \eq{delkernel} and under the background gauge symmetry
\eq{delbarkernel} since it does not depend 
on the background field. 
In turn, neither the gauge fixing term nor the ghost field action
are invariant under \eq{delkernel} or \eq{delbarkernel}. 
Their variation under \eq{delkernel} yields 
\begin{eqnarray}\label{vary1} 
\Der^a(x) S_{\rm gf}\tab = \tab 
{1\ov \xi} D^{ab}_\mu\longbar D{}^{bc}_\mu\longbar D{}^{cd}_\nu 
(A-\bar A)^d_\nu(x) \\\di 
\Der^a(x)  S_{\rm gh}\tab = \tab f^{bdc}\longbar D{}_{\mu}^{ad}\left( c^*_b 
D^{ce}_\mu c_e\right)\ .   
\label{vary2} 
\end{eqnarray} 
However, making use of \eq{gauge}, \eq{ghost} and \eq{delbar} it is 
easy to see, that \eq{vary1} and \eq{vary2} are just  
$-\bar\Der^a S_{\rm gf}$ and $-\bar\Der^a  S_{\rm gh}$ respectively. 
Thus, each term in the action $S[A,c,c^*;\bar A]$ 
is {\it separately} invariant under the
combined transformation $\Der+\bar\Der$. 
This brings us to a key point of the background field formalism.
The invariance of $S[A,c,c^*;\bar A]$ under $\Der+\bar\Der$ implies that
the action $\hat S[A,c,c^*]\equiv S[A,c,c^*;\bar A =A]$
 is invariant under the {\it physical} symmetry 
\eq{delkernel}, $\Der^a \hat S[A,c,c^*]=0$, with $S[A,c,c^*;\bar A]$ 
satisfying the classical 'Ward-Takahashi identity' $\Der^a S=\Der^a 
(S_{\rm gf}+S_{\rm gh})$.  

At quantum level these statements turn into 
gauge invariance of the effective action $\Gamma[A,c,c^*;\bar A=A]$ with 
$\Gamma[A,c,c^*;\bar A]$ satisfying the Ward-Takahashi identity of a 
non-Abelian gauge theory. 
Note that only the combination of both statements 
gives a physical meaning to gauge invariance of $\Gamma[A,c,c^*;\bar A=A]$. 
In the quantised theory where the sources couple only to the
fluctuation field
\begin{eqnarray}
a^a_\mu = A^a_\mu -\bar A^a_\mu\ ,  
\end{eqnarray} 
the resulting theory is evaluated for vanishing
expectation value $\langle a\rangle=0$ (Notice that the gauge fixing
condition~\eq{gauge} only constrains $a^a_\mu$). \step

\noindent
{\bf 3. Wilsonian flows}\\[-1ex]

We now follow the strategy sketched above for the case of a 
coarse-grained effective action along the lines in \cite{Reuter:1994kw}.
Scale-dependent regulator terms for the gauge and the ghost fields,
respectively, are added to the Yang-Mills action \eq{S}, 
\begin{eqnarray}\label{Sk} 
S_k=S+ \Delta S_{k}\ ,\qquad\Delta S_{k}=\Delta S_{k,A}+ \Delta
S_{k,C}\ .
\end{eqnarray}
The new terms are (non-local) operators quadratic in the fields 
and given by 
\begin{eqnarray}        \label{DeltaSA} 
\Delta S_{k,A} &=& \frac{1}{2}
\int_x (A-\bar A)^a_{\mu}\,R_{\!A}{}^{ab}_{\mu\nu}(P_{\!\!\!A}^2)\,
(A-\bar A)^b_{\nu}  
\\                      \label{DeltaSC} 
\Delta S_{k,C} &=&  \int_x c^*_a \,R_{C}^{ab}(P_{\!\!C}^2)\,c_b\ ,   
\end{eqnarray}
where the arguments  $P_{\!\!\!A}^2$ and $P_{\!\!C}^2$ of the
regulator functions are appropriately defined Laplaceans. 
A suitable choice for them in the present context is 
$(P_{\!\!\!A}^2)^{ab}_{\mu\nu}= (-\longbar D{}^2)^{ab}\delta_{\mu\nu}$ and 
$(P_{\!\!C}^2)^{ab}=(-\longbar D{}^2)^{ab}$, which are 
operators with a positive spectrum of eigenvalues. The regulator functions 
$R_A$ and $R_{C}$ depend on both the coarse graining scale $k$, and on the
background field $\bar A^a_\mu$ via $P_{\!\!\!A}^2$ and $P_{\!\!C}^2$. 
A typical example for an exponentially smooth  
regulator function is given by $R(P^2)=P^2/(\exp P^2/k^2 -1)$.

In order to maintain the invariance of $S_k$ under the combined transformation 
$\Der^a+\bar\Der^a$ we have to ensure that both \eq{DeltaSA} and 
\eq{DeltaSC} vanish under $\Der^a+\bar\Der^a$. For the action of $\Der^a$
on the regulator terms we find 
\begin{eqnarray}\label{vary3}
\Der^a(x) \, \Delta S_{k}= 
-D_\mu^{ab}\,R_A{}^{bc}_{\mu\nu}(P_{\!\!\!A}^2)\,(A-\bar A)^c_{\nu} 
- g \int_y c_b^*(y)\left([t^a\delta(x-y),\,R_{\hspace{-.01cm}C}
(P_{\!\!C,y}^2)]\right)^{bc} c_c(y)\ . 
\end{eqnarray} 
To compute how $\bar\Der^a$ operates on \eq{DeltaSA}  
and \eq{DeltaSC}, it is helpful to consider first the action of 
\eq{delbarkernel} on $\longbar D{}^2$. From~(\ref{delcov}) (with $A=\bar A$) 
it follows 
immediately that $\longbar D{}^2$ transforms as a tensor. 
Therefore $P^2$ and $R(P^2)$, which are both functions of
$-\longbar D{}^2$, also transform as tensors under \eq{delbarkernel} 
\begin{eqnarray}\label{trafofR}
\bar\Der^a(x)\,R(P^2_{y})=[t^a\delta(x-y),\,R(P^2_y)] . 
\end{eqnarray}
Using also \eq{delbar} it is straightforward to show that 
\eq{vary3} equals $-\bar\Der^a \left(\Delta S_{k,A}+\Delta S_{k,C}\right)$ 
which establishes the desired property.\footnote{More generally, 
$P_{\!\!\!A}^2$ and $P_{\!\!C}^2$ need not to be of the form as given
in the text.
The required symmetry properties remain unchanged as long as they 
transform as tensors under the background gauge transformation. 
Our choice above is one such example.}

So far, we have restricted the discussion to the classical action \eq{S} 
with the regulator terms \eq{Sk} added. The computation of the effective 
action $\Gamma_k$ follows with the usual procedure.
We introduce sources $(J,\eta,\eta^*)$ for the fields
$(A-\bar A,c,c^*)$ to consider first the Schwinger functional
$W_k \equiv W_k[J_{\mu},\,\eta,\,\eta^*;\, \bar A_\mu]$, given by
\begin{eqnarray}\label{Schwinger}
  \exp W_k = \ds\int\prod_a\left\{\dcal A^a_{\mu}\,\dcal c_a\,\dcal
  c^*_a\right\}\,\exp\left[-S_k
  +\int(J^a_{\mu}(A-\bar A)^a_{\mu}+\eta^*_a c_a - c^*_a\eta_a)\right]\ .
\end{eqnarray}
The effective action $\Gamma_k$ is given by its Legendre transform
\begin{eqnarray}
\Gamma_k[A,c,c^*; \bar A]=
  - W_k[J,\eta,\eta^*;\bar A] - \Delta S_k[A,c,c^*;\bar A]
+\int_x\left(J^a_{\mu}(A-\bar A)^a_{\mu}+\eta^*_a \bar c_a -
  c^*_a\eta_a\right). \label{defofG}
\end{eqnarray}
Here, $\Gamma_k$ is a functional of the expectation values of the
fields  ({\it e.g.} $A-\bar A\equiv -\delta W_k/\delta J$,
etc).\footnote{For simplicity, we do not introduce other names for
these fields because we shall only be concerned with $\Gamma_k$ in the
remaining part of the letter.} 
The flow equation for $\Gamma_k$ has already been
given in \eq{flow}.  

Finally we introduce the effective action $\hat \Gamma_k$ which
corresponds to $\Gamma_k$ evaluated at $\bar A=A$,
\begin{equation}
\hat\Gamma_k[A,c,c^*] \equiv\Gamma_k[A,c,c^*;\, \bar A=A]. 
\end{equation} 
As we shall argue below, this action is
gauge-invariant. Its flow equation is simply given by the
one for $\Gamma_k$ in \eq{flow}, evaluated at $\bar A=A$. It is
important to stress that the flow of $\hat \Gamma_k$, since it depends
on the second functional derivatives of $\Gamma_k$ w.r.t\ the 
dynamical fields (at $\bar A=A$), is
a functional of $\Gamma_k$ and {\it not} a functional of $\hat
\Gamma_k$. This makes it mandatory to study not only the symmetries of
$\hat\Gamma_k$  but also those of $\Gamma_k$.\step

\noindent
{\bf 4. Modified and background field Ward-Takahashi identities}\\[-1ex]

We now turn to a detailed discussion of the Ward-Takahashi identities
related to 
the transformations \eq{delkernel} and \eq{delbarkernel}. Ward-Takahashi 
identities follow from an invariance of the Schwinger functional under gauge 
transformations. In the Wilsonian formalism, these identities are 
modified due to the presence of the regulator terms. 
The identity which follows from considering $\Der^a\Gamma_k$ is denoted 
as the {\it modified} Ward-Takahashi identities (mWI). 
A second identity is derived from the background gauge transformations 
$\bar\Der^a\Gamma_k$, leading to the {\it background field} 
Ward-Takahashi identities (bWI).

Let us first summarise some immediate consequences of the 
invariance of $S_k$ under the action of $\Der^a+\bar\Der^a$. 
It can be read off from the definitions of the Schwinger functional 
\eq{Schwinger} and the effective action \eq{defofG}, that the 
combination $\Der^a+\bar\Der^a$ leaves
the functional $\Gamma_k$ invariant for generic $A$ and $\bar A$, 
\begin{eqnarray}
\left(\Der^a+\bar\Der^a\right)\Gamma_k = 0. 
\label{mWI+bWIex}\end{eqnarray} 
Some comments are in order. Within a Wilsonian approach, the
physical Green's function are approached in the limit $k\to 0$, 
where $\Gamma_k$ approaches the full quantum 
effective action. We have already pointed out that 
the statement of {\it physical} gauge invariance 
corresponds to \eq{mWI+bWIex} at $k=0$, with the fields $A$ 
and $\bar A$ identified, only if $\Gamma_{k=0}$ satisfies the {\it usual} 
Ward-Takahashi identity connected to $\Der^a$. Therefore 
it is necessary to keep track of the action of the transformations $\Der^a$ 
and $\bar\Der^a$ on $\Gamma_k$ separately. \Eq{mWI+bWIex} also implies that 
the effective action $\hat\Gamma[A,c,c^*]$ satisfies 
\begin{eqnarray}
\Der^a \hat\Gamma_k = 0,
\label{physmWI+bWIex}\end{eqnarray}
which for $k=0$ expresses the desired physical gauge invariance.  
Consequently, for $k\neq 0$, physical gauge invariance 
is encoded in the behaviour of 
$\Gamma_k$ under the transformation $\Der^a$. 
This is also evident from the fact that the flow of $\hat\Gamma_k$ 
is a functional of $\Gamma_k$. 

We now give a detailed derivation 
of the related modified Ward-Takahashi identity. We start by 
applying $\Der^a$ to the Schwinger functional $W_k$ \eq{Schwinger}. To be more 
precise, we apply $\Der^a$ to the integration fields variables $A,c,c^*$ 
which leaves $W_k$ invariant since the path integral measure is invariant 
under the action of $\Der^a$ and hence $\Der^a W_k=0$. 
Collecting all terms and making the Legendre transformation to $\Gamma_k$  
yields 
\begin{eqnarray}
 \Der^a \Gamma_k &=& 
\langle\Der^a  \left(S_{\rm gf} +S_{\rm gh}+\Delta S_k 
\right)\rangle_J,  
\label{mWI}\end{eqnarray}
where the expectation value $\langle\cdots\rangle_J$ 
stands for connected Green's functions in the 
external source $(J,\eta,\eta^*)=(\delta_A \Gamma_k,\delta_c\Gamma_k,
\delta_{c^*}\Gamma_k)$. We evaluate the expectation values in \eq{mWI} 
by using \eq{vary1}, \eq{vary2} and \eq{vary3}. After some 
algebra we arrive at 
\begin{eqnarray}
\Der^a(x)\,\Gamma_k = 
\Der^a(x)\,\left(S_{\rm gf}+S_{\rm gh}\right)+ 
\Loop_k^a(x) + \Loop_{R,k}^a(x)\ .
\label{mWIex}\end{eqnarray}
Both $\Loop_k$ and $\Loop_{R,k}$ display loop terms. 
The first term $\Loop_k$ stands for the 
well-known loop contributions to Ward-Takahashi identities in non-Abelian 
gauge theories originating from $\langle \Der^a (S_{\rm gf}+S_{\rm
  gh})\rangle_J$. $\Loop_k$ is given by  
\begin{eqnarray}\nonumber
{\Loop^a_k(x) = g\,\,\Bigg[\,\frac{1}{\xi}\,\,
f^{adb}\,\left(\Big(\longbar D{} \otimes\longbar D{}\Big)^{bc}_{\mu\nu}
G_{\!\!A}{}_{\nu\mu}^{cd}\right)(x,x)
-f^{bdc}\longbar D{}_{\mu,x}^{ad} \left( D^{ce}_\mu
G_{\!C}{}^{eb}\right)(x,x) 
\Bigg]}\hspace{1cm}\\
-\,g^2\longbar D{}_{\mu,x}^{ad} \left( f^{bdc}  \,f^{che}
\Bigl[G_{\!\!A}{}_{\mu\nu}^{hg}{\delta\ov\delta A_\nu^g} +
G_{\!\!AC}{}_\mu^{hg}{\delta\ov\delta c^g}+
G_{\!\!AC^*}{}_{\mu}^{hg}{\delta\ov\delta c^*{}^g}\Bigr] \,
G_{\!C}{}^{eb}\right)(x,x), \label{calL}
\end{eqnarray}
where $(\longbar D{} \otimes\longbar D{})_{\mu\nu}^{ab} =
\longbar D{}_\mu^{ac}\longbar D{}_\nu^{cb}$. We have also introduced 
the propagators $G_{\!\!A}$, $G_{\!C}$, $G_{\!\!AC}$ and
$G_{\!\!AC^*}$, whose inverses are given by 
\begin{eqnarray}
\left.G^{-1}_{k,A}\right.^{ab}_{\mu\nu} \equiv  
 \frac{\de^2\Gamma_k}{\delta A^\mu_a\delta A^\nu_b}
  +R^{ab}_{A,\mu\nu}\ , \tab \quad \quad \tab 
\left.G^{-1}_{k,C}\right.^{ab} \equiv  
\frac{\de^2\Gamma_k}{\delta c_a\delta c_b^*} +R^{ab}_{C}\ ,  
\label{propc}
\\[1ex]
\left.G^{-1}_{\!\!AC}\right.^{ab}_{\mu} \equiv
\frac{\de^2\Gamma_k}{\delta A^\mu_a\delta c_b}\ , \tab \quad \quad \tab  
\left.G^{-1}_{\!\!AC^*}\right.^{ab}_{\mu} \equiv 
\frac{\de^2\Gamma_k}{\delta A^\mu_a\delta c^*_b}\ \label{cross}.  
\end{eqnarray}
The term in the second line in \eq{calL} is a two-loop 
contribution to the Ward-Takahashi identity. To see this more explicitly, let 
us write out the first of its contributions: 
\begin{eqnarray}\label{explicit} 
 \longbar D{}_{\mu,x}^{ad} \left( f^{bdc}f^{che}
\int_y G_{\!\!A}{}_{\mu\nu}^{hg}(x,y){\delta\ov\delta A_\nu^g(y)} 
G_{\!C}{}^{eb}(x,x)\right)\ . 
\end{eqnarray}
Note that $ {\delta\ov\delta A(y)}G_{\!C}(x,x)$ is a loop 
closing at $x$ with a gauge field vertex (at $y$) and $G_{\!\!A}(x,y)$ 
is a line emanating at $x$ and connecting to the vertex at $y$
(see figure). \\[-14ex]
  \begin{figure}[htbp]
    \begin{center}
\begin{picture}(50,150)
      \leavevmode
\put(-34,45){\Large $\mu,d$}
      \psfig{file=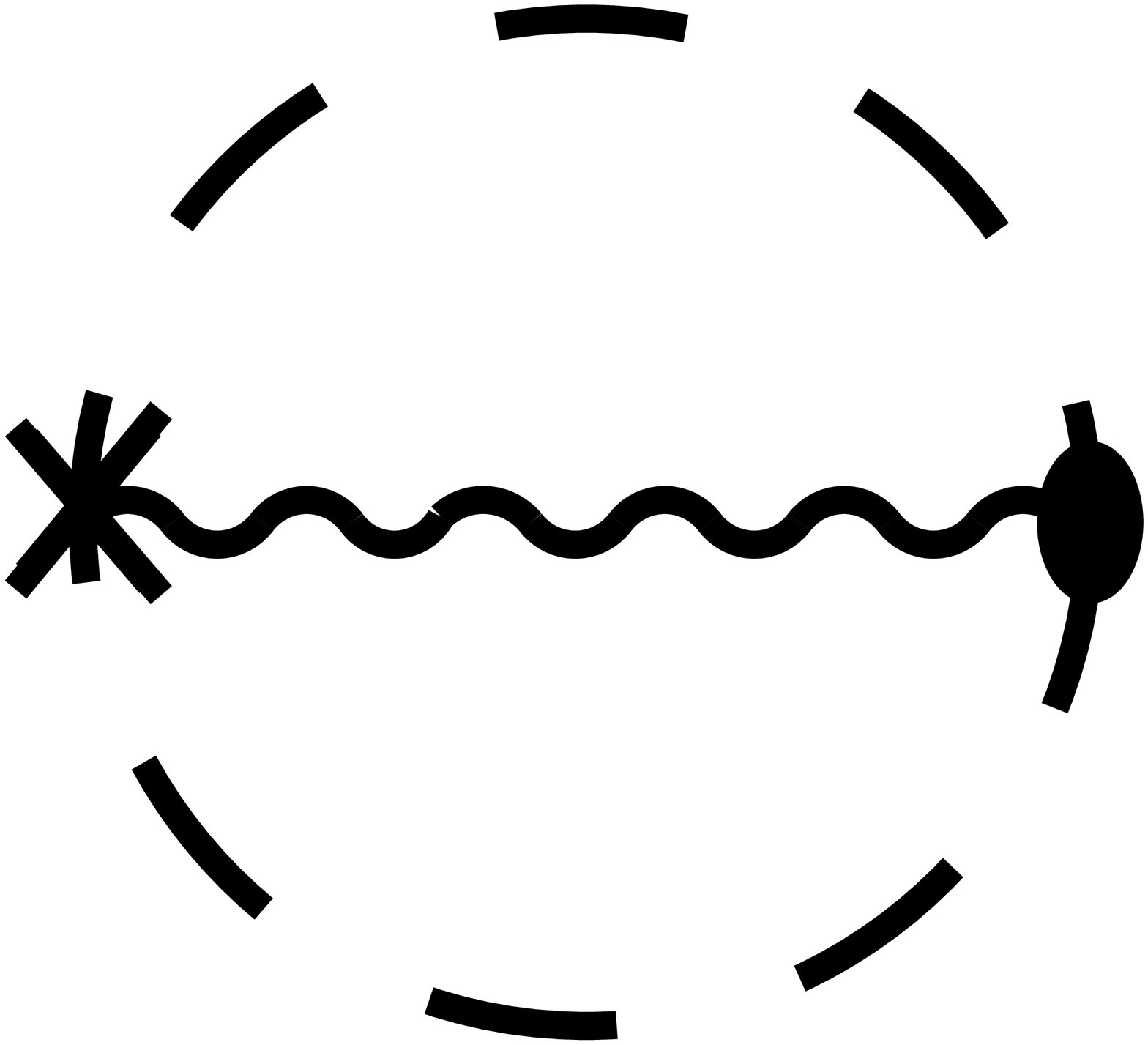,width=.2\hsize,height=.15\vsize}      
\end{picture}
      \vskip0cm
      \begin{minipage}{12.5cm}
        \begin{center}
          \label{Feynman}\textsl{\footnotesize Feynman-diagram for the
            expression in braces in \eq{explicit}}
\end{center}
      \end{minipage}
    \end{center}
  \end{figure}
The term $\Loop_{R,k}$ in \eq{mWIex} comprises 
the loop contribution of $\langle \Der^a \Delta S_k\rangle_J$ coming directly 
from the coarse graining and it is given by
\begin{eqnarray}
\Loop_{R,k}^a(x) = g \tr_{\rm ad}\,t^a\,\left(\left[ 
G_{\!\!A}^{\mu\nu},\, R_{\!A}^{\nu\mu}(P_{\!\!\!A}^2)\right](x,x)
- \left[G_{\!C},\,
R^{}_{\hspace{-.01cm}C}(P_{\!\!C}^2)\right](x,x)\right)\ .
\label{calLR} \end{eqnarray}
This term clearly disappears in the limit $k\to 0$, $\Loop_{R,0}\equiv 0$. 

The identity expressed by \eq{mWIex} is the modified Ward-Takahashi 
identities where the contribution from the coarse graining is 
contained in $\Loop_{R,k}$.
It follows that the mWI \eq{mWIex} turns into the usual WI for $k=0$:
\begin{eqnarray}
\Der^a\Gamma = \Loop^a_0. 
\label{WI}\end{eqnarray}
It is left to cast $\bar\Der^a\Gamma_k$ into an explicit form. 
Starting by applying $\bar\Der^a$ to 
$W_k[J,\eta,\bar\eta;\bar A]$ leads to $\bar\Der^a W_k= 
-\longbar D{}_\mu^{ab} J^b_\mu-\langle 
\bar \Der^a  \left(S_{\rm gf} +S_{\rm gh}+\Delta S_k
\right)\rangle_J$. It follows from \eq{defofG} 
that 
\begin{eqnarray}\label{expectG}
\bar\Der^a (\Gamma_k+\Delta S_k)=\langle 
\bar \Der^a  \left(S_{\rm gf} +S_{\rm gh}+\Delta S_k
\right)\rangle_J. 
\end{eqnarray} 
Using \eq{vary1}, \eq{vary2}, 
\eq{vary3} and \eq{trafofR} this last equation takes the form
\begin{eqnarray}
 \bar\Der^a\Gamma_k = 
\bar\Der^a(S_{\rm gf}+S_{\rm gh})-(\Loop^a_k+\Loop^a_{R,k}). 
\label{bWIex}\end{eqnarray}
Eq.~(\ref{mWI+bWIex}) follows immediately from this identity and the
mWI~(\ref{mWIex}).

We close this section with a comment on the finiteness of 
\eq{mWIex} and \eq{bWIex}. The quantum corrections in $\Loop^a_k$ are
familiar as they appear already in the usual WI.  In perturbation
theory, these terms require an additional UV regularisation and
renormalisation.
In the present Wilsonian framework, however, we are dealing
with UV {\it finite} quantities.
The flow of the effective action $\Gamma_k$ starts with a finite
initial condition at $k=\Lambda$, $\Gamma_{\Lambda}$.
The finiteness of $\Loop^a_k$ then follows from the observation that 
the flow (which is IR {\it and} UV finite) cannot generate
UV divergences.
Therefore, in contrast to perturbation theory, no additional renormalisation 
is needed to make them finite, which is one of the key
advantages of the ERG approach.\step

\noindent
{\bf 5. Symmetries of the flow and physical gauge invariance}\\[-1ex]

The implementation of coarse graining modifies the gauge
symmetry of the theory as we have discussed. 
At the formal level it is clear that the original 
symmetry is restored when the coarse graining scale is removed (see
also \eq{WI}). A more delicate problem is to guarantee 
that this also happens at the level of the solution to the flow equation.

To understand how gauge invariance is encoded throughout the flow, 
it is pivotal to also study the action of the symmetry transformations 
on $\partial_t\Gamma_k$ (see \eq{flow}). 
Firstly we derive how the combined transformation 
$\Der^a+\bar\Der^a$ acts on $\partial_t\Gamma$ where we only 
want to argue at the level of the flow equation. 
The flow equation \eq{flow} functionally depends on second derivatives of 
$\Gamma_k$ w.r.t.\ fields $A,c,c^*$ and on $R,\partial_t R$. 
Hence, we are interested on the action of $\Der^a+\bar\Der^a$ on 
these quantities. We note that 
\begin{eqnarray} \nonumber
(\Der^a+\bar\Der^a)(x) {\delta^2 \Gamma_k\ov \delta A_b^\mu  
\delta A_c^\nu} &=&\di   
\Bigl(\Bigl[\delta_x t^a,\,{\delta^2 \Gamma_k\ov \delta A_\mu  
\delta A_\nu}\Bigr]\Bigr)^{bc}, \\\di 
(\Der^a+\bar\Der^a)(x)
\,{\delta^2 \Gamma_k\ov \delta c_b\delta c^*_c} & =&\di   
\Bigl(\Bigl[\delta_x t^a,\,{\delta^2 \Gamma_k\ov \delta c\delta c^*}
\Bigr]\Bigr)^{bc} 
\label{trafofdG}\end{eqnarray}
and similar identities for mixed derivatives. Here we have used \eq{mWI+bWIex} 
and the 
commutator of two derivatives w.r.t. the fields $A,c,c^*$ and 
$\Der^a$. For the sake of simplicity we have 
introduced the short hand notation 
$[\delta_x,\,\CO](y,z)=\delta(y-x)\CO(y,z)-\CO(y,z)\delta(z-x)$. This 
facilitates the following conclusion. 
\Eq{trafofdG} states that second derivatives of 
$\Gamma_k$ w.r.t.\ the fields $A,c,c^*$ transform as tensors under 
$\Der^a+\bar\Der^a$. Together with \eq{trafofR} this implies that the 
propagators $G_{\!\!A}$ and $G_{\!C}$ transform as tensors: 
\begin{eqnarray}\label{trafofprops} 
(\Der^a+\bar\Der^a)(x)G_{\!\!A}=[t^a\delta_x,\,G_{\!\!A}], \qquad 
(\Der^a+\bar\Der^a)G_{\!C}=[t^a\delta_x,\,G_{\!C}]. 
\end{eqnarray}
With \eq{trafofR} and \eq{trafofprops} we conclude 
\begin{eqnarray} 
(\Der^a+\bar\Der^a) \partial_t\Gamma_k=0. 
\label{d_Bflow}\end{eqnarray}
This implies that $\Der^a\partial_t\hat\Gamma_k=0$. 
Note that the only input for \eq{d_Bflow} was the invariance of $\Gamma_k$. 
Thus, if the initial effective action $\Gamma_{\Lambda}$ is invariant under 
$\Der^a+\bar\Der^a$ 
it follows that the full effective action $\Ga_0$ satisfies 
$(\Der^a+\bar\Der^a)\Ga_0=0$. 
In other words, \eq{mWI+bWIex} and \eq{d_Bflow} are 
the proof that flow and $(\Der^a+\bar\Der^a)$ commute. Moreover, 
$\Gamma_0$ satisfies the usual WI \eq{WI}. 
This means that we can follow the line of arguments of the background 
field formalism as explained in the second section. 
Thus we conclude that $\Der^a \hat\Gamma_0=0$  displays 
physical gauge invariance.  

Now we continue with a remark on the consistency of the mWI 
\eq{mWIex} with the flow. As for other formulations of Wilsonian flows in 
gauge theories \cite{Ellwanger:1994iz,D'Attanasio:1996jd,Litim:1998qi,Litim:1998nf}, 
the flow of the modified Ward-Takahashi identity is proportional to
the mWI itself. Very schematically this identity has the form 
\begin{eqnarray}\label{dmWI0} 
\Bigl(\partial_t-{\cal O}\Bigr)         
(\Der^a \Gamma_k-\Der^a \left(S_{\rm gf}+S_{\rm gh}\right)- 
\Loop^a_k - \Loop_{R,k}^a)=0, 
\end{eqnarray}  
where $\cal O$ does only depend on $\Gamma_k$ and derivatives of the field. 
An explicit expression for $\cal O$ in \eq{dmWI0} 
has been given in \cite{Litim:1998nf}. 
Eq.~\eq{dmWI0} states that if the effective action $\Gamma_k$ satisfies 
the mWI at some (initial) scale $k=\Lambda$, then $\Gamma_k$ 
 {\it automatically} satisfies the mWI for all scales $k$, provided it is 
obtained from integrating the flow equation. Hence, $\Gamma_0$ 
satisfies the usual Ward-Takahashi identity.

We also like to briefly discuss the connection between gauge invariance
in the present approach and BRST-invariance in the BRST formalism.
An analogous treatment within a BRST formulation has been
given \cite{Ellwanger:1994iz,Bonini:1994sj,D'Attanasio:1996jd} where
the information about the gauge symmetry is carried by a modified
BRST-identity. This identity reduces to the mWI \eq{mWIex} by
integrating out the ghosts and putting the BRST charges to zero. Its
advantage in the standard perturbative approach is that BRST
invariance leads to a bilinear equation in derivatives of the
effective action (the well-known master equation).  In the presence of
a coarse graining term, however, this identity receives an additional
term which contains the propagator derived from the effective action,
thus spoiling the bilinear structure. This additional term has the
same form as the loop terms already present in Ward-Takahashi
identities for non-linear symmetries. Therefore, we see no advantage
in studying BRST invariance rather than the usual Ward-Takahashi
identities.

In summary, we have established the complete set of
equations relevant for the control of gauge invariance
within the present approach.
In particular, it has been shown that there is no additional fine-tuning
condition, despite the presence of a background field.  These results
can be used for further interesting applications in the Wilsonian
approach to non-Abelian gauge theories. The formalism is
well-suited for analytic calculations.  For example a consistent
calculation of 2-loop quantities can be done analytically \cite{J}.
Note that the introduction of a background field to the cut-off terms
is not restricted to the background field gauge discussed here.  In
axial gauges a similar procedure can be used to obtain a gauge
invariant effective action. Here analytic calculations of the full
effective action (in some approximation) are accessible
\cite{DJ}.\\

\noindent
{\bf Acknowledgements}\\[-2ex]

It is a pleasure to thank 
C.~Wetterich for discussions. FF and JMP thank the
Institut f\" ur 
Theoretische Physik, Heidelberg and DFL the Dublin Institute of 
Advanced Studies for their kind hospitality during different
stages of the work.\\[-2ex]


\noindent

\begin{thebibliography}{99}
\def\BOOK#1#2#3#4{#1, {\sc } #3, #4}
\def\PRA#1#2#3#4#5{ #1,\,\,{\it }\,Phys.\,Rev.\,{\bf A#3}\,(19#4)\,#5}
\def\PRB#1#2#3#4#5{#1,{\it }\,Phys.\,Rev.\,{\bf B#3}\,(19#4)\,#5}
\def\PRL#1#2#3#4#5{#1,\,\,{\it }\,Phys.\,Rev.\,Lett.\,{\bf #3} (19#4) #5}
\def\PRC#1#2#3#4#5{#1,\,\,{\it }\,Phys.\,Rev.\,{\bf C#3}\,(19#4)\,#5}
\def\PRD#1#2#3#4#5{#1,{\it }\,Phys.\,Rev.\,{\bf D#3}\,(19#4)\,#5}
\def\PRE#1#2#3#4#5{#1,\,\,{\it }\,Phys.\,Rev.\,{\bf E\,#3}\,(19#4)\,#5}
\def\PRep#1#2#3#4#5{#1,{\it }\,Phys.\,Rep.\,{\bf  #3}\,(19#4)\,#5}
\def\NPB#1#2#3#4#5{#1,{\it }\,Nucl.\,Phys.\,{\bf B#3}\,(19#4)\,#5}
\def\PLB#1#2#3#4#5{#1,{\it }\,Phys.\,Lett.\,{\bf B#3}\,(19#4)\,#5}
\def\PTP#1#2#3#4#5{#1,\,\,{\it }\,Prog.\,Theor.\,Phys.\,{\bf B#3}\,(19#4)\,#5}
\def\SSC#1#2#3#4#5{#1,\,\,{\it }\,Solid\,State\,Comm.\,{\bf #3} (19#4) #5}
\def\EPL#1#2#3#4#5{#1,\,\,{\it }\,Europhys. Lett.~{\bf #3}\,(19#4)\,#5}
\def\JCP#1#2#3#4#5{#1,\,\,{\it }\,J.\,Phys.\,(Paris) {\bf  #3} (19#4) #5}
\def\JPA#1#2#3#4#5{#1,\,\,{\it }\,J.\,Phys.\,{\bf A#3}\,(19#4)\,#5}
\def\JPB#1#2#3#4#5{#1,\,\,{\it }\,J.\,Phys.\,{\bf B#3}\,(19#4)\,#5}
\def\JPC#1#2#3#4#5{#1,\,\,{\it }\,J.\,Phys.\,{\bf C#3}\,(19#4)\,#5}
\def\ZPC#1#2#3#4#5{#1,{\it }\,Z.\,Phys.\,{\bf C#3}\,(19#4)\,#5}
\def\JETP#1#2#3#4#5{#1,\,\,{\it }\,Soviet\,Physics\,JETP\,Lett.\,{\bf #3}\,(19#4)\,#5}
\def\MPLA#1#2#3#4#5{#1,{\it }\,Mod.\,Phys.\,Lett.\,{\bf A#3}\,(19#4)\,#5}
\def\PA#1#2#3#4#5{#1,\,\,{\it }\,Physica\,{\bf A#3}\,(19#4)\,#5}
\def\PS#1#2#3#4#5{#1,\,\,{\it }\,Physics\.{\bf #3}\,(19#4)\,#5}
\def\AP#1#2#3#4#5{#1,\,\,{\it }\,Ann.\,Phys.\,{\bf #3}\,(19#4)\,#5}
\def\IJMPA#1#2#3#4#5{#1,{\it }\,Int.\,J.\,Mod.\,Phys.\,{\bf A#3}\,(19#4)\,#5}
\def\LNC#1#2#3#4#5{#1,\,\,{\it }\,Lett.\,Nuevo\,Cimento\,{\bf #3}\,(19#4)\,#5}
\def\PPR#1#2#3{#1,\,\,{\it }\,Preprint\,#3}
\def\and#1#2#3{{\bf #1}\,(19#2)\,#3}

\bibitem{Polchinski:1984gv}
J.~Polchinski,
Nucl.\ Phys.\  {\bf B231} (1984) 269.

\bibitem{Reuter:1994kw}
M.~Reuter and C.~Wetterich,
Nucl.\ Phys.\  {\bf B417} (1994) 181.

\bibitem{Ellwanger:1994iz}
U.~Ellwanger,
Phys.\ Lett.\  {\bf B335} (1994) 364
[hep-th/9402077].

\bibitem{Ellwanger:1998wy}
U.~Ellwanger, M.~Hirsch and A.~Weber,
Z.\ Phys.\  {\bf C69} (1996) 687
[hep-th/9506019]; Eur.\ Phys.\ J.\  {\bf C1} (1998) 563
[hep-ph/9606468].

\bibitem{Bonini:1994sj}
M.\,Bonini, M.\,D'Attanasio and G.\,Marchesini,
Nucl.\ Phys.\ {\bf B421} (1994) 429
[hep-th/ 9312114];\ 
Phys.\ Lett.\  {\bf B346} (1995) 87;
[hep-th/9412195];\  
Nucl.\ Phys.\  {\bf B437} (1995) 163
[hep-th/9410138].

\bibitem{Freire:1996db}
F.~Freire and C.~Wetterich,
Phys.\ Lett.\  {\bf B380} (1996) 337
[hep-th/9601081].

\bibitem{D'Attanasio:1996jd}
M.~D'Attanasio and T.R.~Morris,
Phys.\ Lett.\  {\bf B378} (1996) 213
[hep-th/9602156].


\bibitem{Litim:1998qi}
D.~F.~Litim and J.~M.~Pawlowski,
Phys.\ Lett.\  {\bf B435} (1998) 181
[hep-th/9802064];\ 
Nucl.\ Phys.\ Proc.\ Suppl.\  {\bf 74} (1999) 325
[hep-th/9809020];\ 
Nucl.\ Phys.\ Proc.\ Suppl.\  {\bf 74} (1999) 329
[hep-th/9809023].

\bibitem{Morris:2000px}
T.~R.~Morris,
Nucl.\ Phys.\  {\bf B573} (2000) 97
[hep-th/9910058]; 
hep-th/0006064.

\bibitem{Litim:1998nf}D.F.~Litim and J.M.~Pawlowski, Proceedings of the 
Workshop on the ERG in Faro, Portugal, Sep98, published in 
World Scientific [hep-th/9901063].

\bibitem{Abbott:1981hw}
L.F.~Abbott,
Nucl.\ Phys.\  {\bf B185} (1981) 189.

\bibitem{J}J.M.~Pawlowski, in preparation. 

\bibitem{DJ}
D.F.~Litim and J.M.~Pawlowski, in preparation.
\end{thebibliography}
\end{document}